\documentclass[twocolumn]{aa}
\usepackage{graphicx}
\usepackage{txfonts}

\begin{document}

   \title{Effects of line-blocking \\ 
		on the non-LTE {Fe}~{I} spectral line formation}

   \author{R. Collet \inst{1}
          \and
          M. Asplund \inst{2}
          \and
	  F. Th\'evenin \inst{3}}

   \offprints{R. Collet,\\	\email{remo@astro.uu.se}}

   \institute{Department of Astronomy and Space Physics, 
        Uppsala University, BOX 515, SE--751 20 Uppsala, Sweden
        \and
	Research School of Astronomy \& Astrophysics,
	Mount Stromlo Observatory, Cotter Road, Weston ACT 2611, Australia
	\and
	Observatoire de la C\^ote d'Azur,
	B.P. 4229, F--06304 Nice Cedex 4, France
	}

   \date{Received 4 May 2005 ; accepted 11 July 2005}

   \abstract{The effects of background line opacity (line-blocking) in
   statistical equilibrium calculations for \element{Fe} in late-type stellar 
   atmospheres have been investigated using an extensive and
   up-to-date model atom with radiative data primarily from the {\sc iron} Project. 
   The background metal line opacities have been computed 
   using data from the {\sc marcs} stellar model atmospheres. 
   While accounting for this line opacity is important at solar metallicity,
   the differences between calculations including and excluding line-blocking at
   low metallicity are insignificant for the non-local thermodynamic equilibrium 
   (non-LTE) abundance corrections for \ion{Fe}{i} lines. 
   The line-blocking has no impact on the non-LTE effects of \ion{Fe}{ii} lines. 
   The dominant uncertainty in \element{Fe} non-LTE calculations for metal-poor stars
   is still the treatment of the inelastic \ion{H}{i} collisions,
   which have here been included using scaling factors to the classical 
   Drawin formalism, 
   and whether or not thermalisation of the high 
   \ion{Fe}{i} levels to \ion{Fe}{ii} ground state should be enforced.
   Without such thermalisation, the \ion{Fe}{i} non-LTE abundance corrections
   are substantial in metal-poor stars: about $0.3$~dex with efficient (i.e. Drawin-like)
   \ion{H}{i} collisions and $\la0.5$~dex without. Without both thermalisation and
   \ion{H}{i} collisions, even \ion{Fe}{ii} lines show significant non-LTE effects
   in such stars.

	}
\authorrunning
\titlerunning
\maketitle

\section{Introduction}
Accurate knowledge of stellar Fe abundances is crucial for our 
understanding of stellar evolution as well as for the study of 
cosmic chemical evolution.
Indeed the abundance of iron is normally taken as a proxy for 
the stellar metal content.
Recent work has been devoted to quantify possible departures 
from local thermodynamic equilibrium (LTE) in the formation
of Fe I spectral lines but without reaching
any consensus to the expected magnitude of the non-LTE effects
in late-type stars (e.g. Gratton et al. \cite{gratton}; 
Th\'evenin \& Idiart \cite{thevenin}; Shchukina \& Trujillo Bueno \cite{shchukina}; 
Gratton et al \cite{gratton2}; Gehren et al. \cite{gehren}, \cite{gehren2};
Korn et al. \cite{korn}; Shchukina et al. \cite{shchukina2};
see discussion and further references in Asplund \cite{asplund2}).
The different results are particularly striking at low metallicity
where some authors find very large non-LTE abundance corrections 
($>+0.3$\,dex) while others find negligible effects ($<+0.1$\,dex). 
It is embarrassing that such large discrepancies
still exist for the most commonly studied element. It certainly raises
concern over the astronomical interpretations not only for Fe but for other
elements as well. 
 
Since the main non-LTE physical mechanism for Fe -- over-ionisation --
feeds on the UV radiation field it is paramount to include
the effects of line-blocking in the computations.
This study has been motivated by the fact that
many of the above-mentioned non-LTE calculations have been performed
without properly including this line-blocking. 
For example, background line opacities were not accounted for in detail
in the influential non-LTE study of Th\'evenin \& Idiart (1999). 
Instead artificial multiplication factors  were applied 
to the continuous opacities in their calculations in order to simulate the line haze in the UV. 
Could the large non-LTE abundance corrections they found be significantly
over-estimated because of this shortcoming?

Here we attempt to quantify the effect of 
line-blocking on the determination of non-LTE abundance corrections
for \ion{Fe}{i} lines.
In order to achieve this, we have modified the widely used
statistical equilibrium code {\sc multi} (Carlsson 1986) and as such
our work will also be of interest for future 
non-LTE line formation studies for other elements. 
In view of the remaining uncertainties attached to the treatment
of inelastic \ion{H}{i} collisions (e.g. Asplund \cite{asplund2}), our focus 
in this initial investigation is not
on the exact values of the non-LTE abundance corrections but on the
differential impact of line-blocking in
late-type stars of different metallicity.
We defer a comprehensive investigation of the non-LTE line formation
for a large grid of stellar model atmospheres to a future
study. We also intend to return to the problem of Fe line formation
using 3D model atmospheres.  

\section{Methods}
\subsection{The model atom}
\subsubsection{Energy levels and radiative data}
For the non-LTE \ion{Fe}{i} spectral line formation calculations in this 
paper we adopted an updated version of the \element{Fe} model atom
by Th\'evenin \& Idiart (\cite{thevenin}).
At the present time the \ion{Fe}{i} and \ion{Fe}{ii} systems in the 
National Institute of Standards and Technology ({\sc NIST}) 
database\footnote{\texttt{http://physics.nist.gov/PhysRefData/ASD/}}
consist of 493 and 578 identified levels respectively (Sugar \& Corliss \cite{sugar}).
The laboratory analysis by Nave et al. (\cite{nave}) has further
extended the information on the  \ion{Fe}{i} system by bringing the
number of identified energy levels to 846 and the number of identified 
lines to nearly 9\,800.
In recent years moreover the ongoing international collaboration known as the
{\sc ferrum} Project (Li et al. \cite{li}; Johansson et al. \cite{johansson}) 
has aimed at expanding and improving the database of oscillator strengths and
transition probabilities of \ion{Fe}{ii} lines of astrophysical significance
by means of both laboratory measurements and theoretical calculations.
However, as the {\sc ferrum} Project has only addressed a selection of  
\ion{Fe}{ii} bound-bound transitions at this stage and is not yet complete
we relied on the data provided by {\sc NIST} to construct the \ion{Fe}{ii} sub-system
in our work.
In practice data storage and computing time impose 
limitations to the complexity of the actual model atom chosen
for the calculations.
We therefore restricted our model to 334 levels of \ion{Fe}{i} from the 
{\sc NIST} database up to an excitation potential of $6.91$~eV 
(about 1~eV below the first ionisation edge),
189 levels of \ion{Fe}{ii} up to $16.5$~eV above the \ion{Fe}{i}
ground state, and the ground level of \ion{Fe}{iii}.
In total 3\,466 bound-bound radiative transitions between
\ion{Fe}{i} levels and 3\,440 between \ion{Fe}{ii} levels
were considered, together with the photo-ionisation
cross-sections for 523 bound-free transitions.

Spectral line profiles were sampled for typically 70 wavelength points
and approximated with the standard Voigt function.
 The $gf$-values for \ion{Fe}{i} lines were taken from
Nave et al. (\cite{nave}) and Kurucz \& Bell (\cite{kurucz}) databases,
while for the oscillator strengths of \ion{Fe}{ii} lines we 
referred to the same sources as Th\'evenin \& Idiart (\cite{thevenin}).
The radiative data for the photo-ionisation cross-sections
came from the {\sc iron} Project (Bautista \cite{bautista}).
The detailed bound-free cross-sections were smoothed whenever
possible to reduce the number of wavelength points in the
calculations and therefore decrease the computing time. 
When strong resonances were present near the photo-ionisation
edges we kept the regular frequency sampling.
We tabulated the cross-sections for bound-free transitions from 
\ion{Fe}{i} levels for typically about 100 wavelength points 
on average up to 300 wavelength points when high-resolution 
was necessary. Cross-sections for photo-ionisations from \ion{Fe}{ii} levels
were tabulated instead for about 30 wavelength points on average.

\subsubsection{Collisional data}

\begin{table}[tbp]
      	\caption{Stellar parameters for which the {\sc marcs}
		model atmospheres used in this paper were computed. 
		The iron abundance is expressed in the notation 
      		$[\mathrm{Fe/H}]\equiv\log(n_{\element{Fe}}/n_{\element{H}})
		-\log(n_{\element{Fe}}/n_{\element{H}})_{\sun}$.
      		A solar iron abundance $\log\epsilon_{\element{Fe},\sun}=7.50$
		is assumed. }
		
	\label{marcsparam}
	\begin{tabular}{p{0.45\linewidth}ccr}
		\hline
		\hline
        	\noalign{\smallskip}
		Object & $T_\mathrm{eff} / [\mathrm{K}]$ 
		& $\log g$ & $[\mathrm{Fe/H}]_\mathrm{LTE}$\\
		\noalign{\smallskip}
		\hline
		\noalign{\smallskip}
		\object{Sun} &  $5\,780$ &  $4.44$ & $0.0$ \\
		\object{Procyon}   &  $6\,500$ &  $4.00$ & $0.0$ \\
		\object{HD~140283} &  $5\,700$ &  $3.70$ & $-2.5$ \\
		\object{G64-12}    &  $6\,400$ &  $4.10$ & $-3.3$ \\
		\noalign{\smallskip}
		\hline
       \end{tabular}
\end{table}

\begin{figure}[thp]
   	\resizebox{\hsize}{!}{\includegraphics{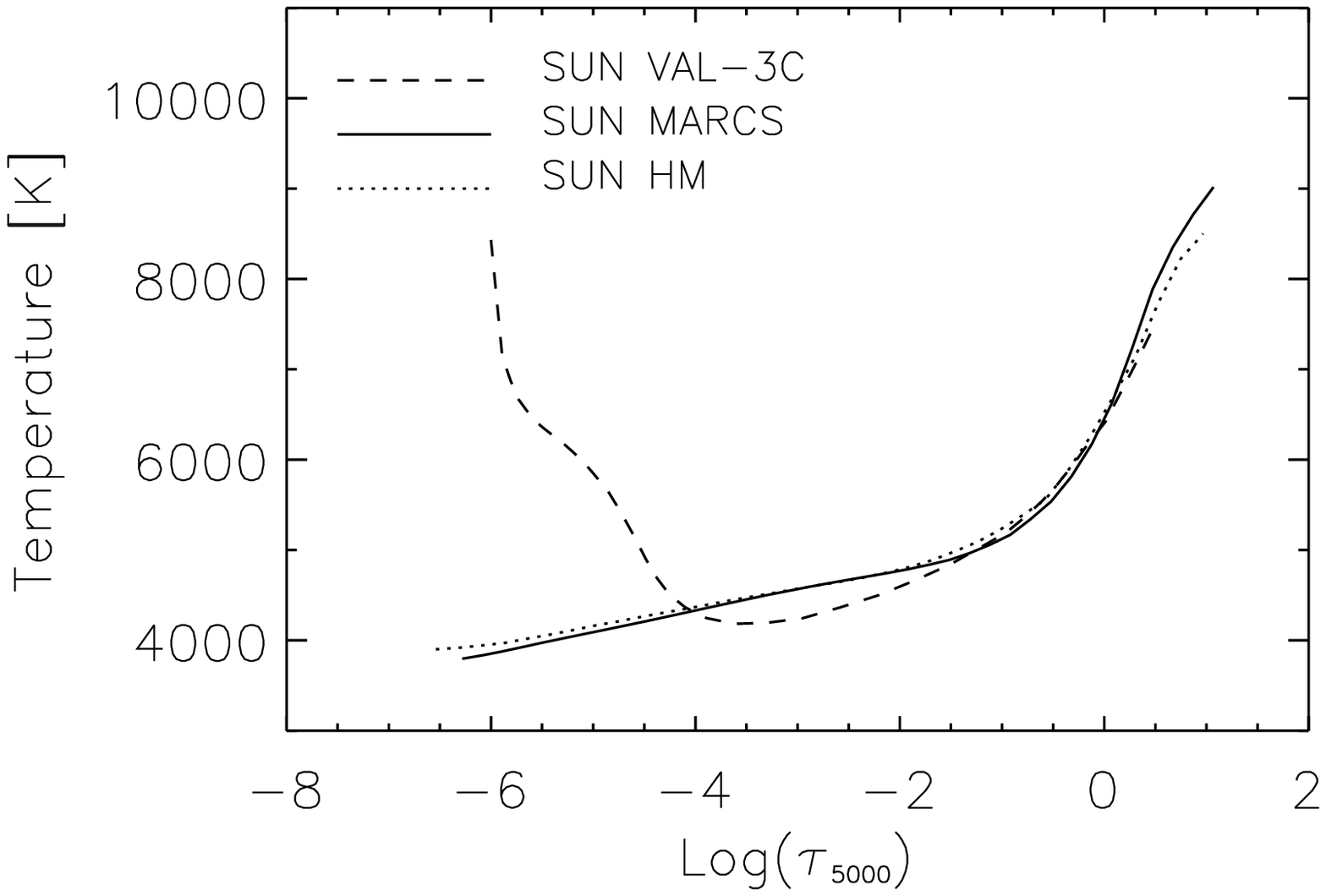} }
   	\resizebox{\hsize}{!}{\includegraphics{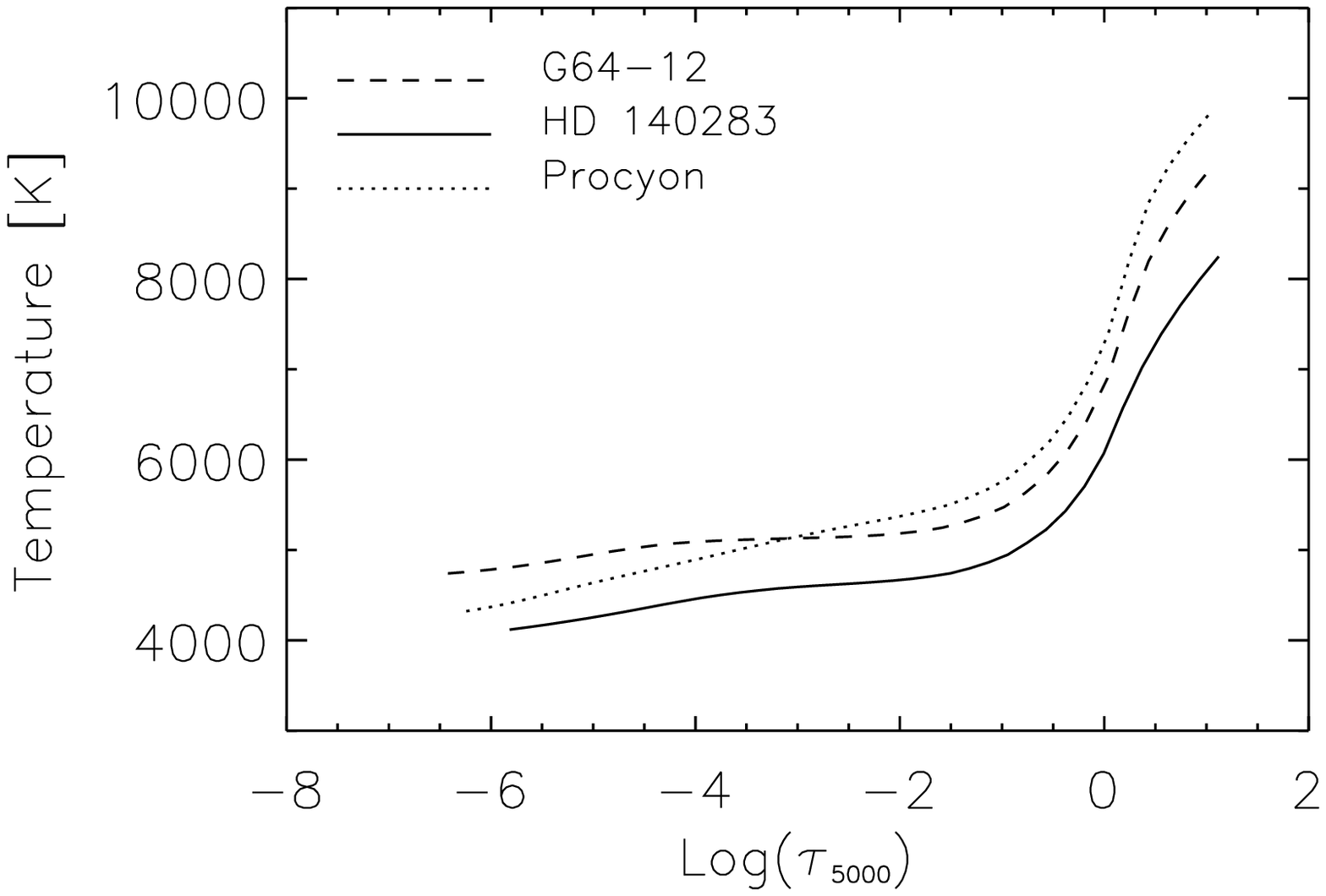} }
      	\caption{Temperature stratification versus standard optical
      		depth at $\lambda=5000$~{\AA} for the three models solar
      		atmosphere adopted in this paper (upper panel) and for 
      		the {\sc marcs} model atmospheres of \object{Procyon}, 
      	\object{HD~140283} and \object{G64-12} (lower panel). }
   	\label{marcsatm}
\end{figure}

All levels in our \element{Fe} model atom were coupled 
with one another via collisional excitation and ionisation by electrons and
by neutral hydrogen atoms, as these types of collisions
are believed to be the most relevant in the atmospheres of cool stars. 
Collisional excitations by electrons were implemented
using van Regemorter's formula (\cite{regemorter})
while cross-sections for ionisation by electron impact
were computed following Cox (\cite{allen}).
All oscillator strengths $f$ were set to a minimum value of $10^{-3}$
in the formulas for bound-bound collisions with electrons and hydrogen atoms.
For the latter type of collisions we adopted the 
approximate description proposed by Drawin (\cite{drawin1}, 
\cite{drawin2}), as done by Steenbock \& Holweger (\cite{steenbock}).
However, both laboratory measurements and 
quantum-mechanical calculations for simple atoms
indicate that the Drawin's recipe fails to reproduce
the right order of magnitude for the cross-sections
of \ion{H}{i} collisions: in the cases for which 
comparison with experimental data or theoretical
calculations is possible, cross-sections computed with the
aid of Drawin's formula appear to be overestimated by
factors of $10-10^6$ depending on the species and
transition (Fleck et al. \cite{fleck}; Lambert \cite{lambert};
Belyaev et al. \cite{belyaev}; Belyaev \& Barklem \cite{belyaev2}; Barklem et
al. \cite{barklem}).

On the other hand, some recent stellar spectroscopic studies suggest
that the cross-sections 
of \ion{H}{i} collisions with \element{Fe} atoms might 
actually be reasonably well described by Drawin's recipe.
Korn et al. (\cite{korn}) found that ionisation equilibrium between
\ion{Fe}{i} and \ion{Fe}{ii} in cool metal-poor stars
is matched consistently with surface gravities derived from
{\sc hipparcos} parallaxes when the Drawinian collisional cross-sections 
are scaled by a factor $S_{\mathrm{H}} \approx 3$ 
using their particular model dependent $T_\mathrm{eff}$ scale (based on
H lines).
Gratton et al. (\cite{gratton}) calibrated \ion{H}{i} collisions
with \element{Fe} by using RR Lyrae variables at their minimum light
and adopted an even larger scaling factor $S_{\mathrm{H}} \approx 30$.
The observational evidence is, however, not unambiguous as some
studies have found the presence of large differences between
\ion{Fe}{i} and \ion{Fe}{ii} based LTE abundances in halo stars
(e.g. Nissen et al. \cite{nissen}; Fuhrmann \cite{fuhrmann}; 
Allende Prieto et al. \cite{allende}) which 
might indicate that the \ion{H}{i} collisions be inefficient compared with 
Drawin's recipe.
Other observational studies suggest vanishing or very small differences
between the \ion{Fe}{i} and \ion{Fe}{ii} based abundances (e.g. Gratton et al. \cite{gratton2};
Kraft \& Ivans \cite {kraft}).
The efficiency of \ion{H}{i} collisions is clearly an open issue which 
requires further work both observationally and, most importantly,
through detailed atomic physics calculations. 

As a reliable and comprehensive description of neutral hydrogen collisions 
with iron atoms is not available at this stage,
we treated the efficiency of \ion{H}{i} collisions \textit{{\`a} la} Drawin
as a free parameter throughout our study.
 In particular, considering the large uncertainty on the actual 
cross-sections of \ion{H}{i} collisions, we decided to apply 
multiplication factors $S_{\mathrm{H}}=0.001$ and $S_{\mathrm{H}}=1$
to Drawin's formula and compared the resulting effects 
on the non-LTE calculations in the two cases.
This approach is justified since the main focus of the present paper
is to investigate the effects of line-blocking 
rather than providing the
final word on the size of the \element{Fe} non-LTE abundance corrections for 
individual stars.

\subsection{The model atmospheres}
As the purpose of our study was to investigate the effects of line-blocking 
on the non-LTE formation of \ion{Fe}{i} spectral lines, 
we have not performed non-LTE calculations for a large
grid of models. Instead we have selected a sample of four cool stars (the \object{Sun},
\object{Procyon}, \object{HD~140283} and \object{G64-12}), covering
the most relevant parameter space and performed the necessary
test calculations.

 The analysis was carried out using {\sc marcs} line-blanketed,
plane-parallel model atmospheres (Asplund et al. \cite{asplund})
of the four objects, computed for the set of stellar parameters 
indicated in Table \ref{marcsparam} and a micro-turbulence 
$\xi=1.0$~km\,s$^{-1}$. 
We kept these effective temperatures and surface gravities constant
throughout the analysis, as the main purpose of our investigation 
is to study differentially the effect of the inclusion and exclusion
of line-blocking on the derivation of non-LTE corrections to the iron 
abundance with respect  to the LTE case and not to fine-tune stellar
parameters. 
We note that our conclusions are immune to reasonable changes of the
adopted stellar parameters. In particular the computed \element{Fe}
non-LTE abundance corrections are affected by about $0.01$~dex or less
when a micro-turbulence $\xi=2.0$~km\,s$^{-1}$ is used instead.

Finally, we adopted for the Sun the  solar chemical 
composition by Grevesse \& Sauval (\cite{grevesse}), i.e. assuming 
an iron abundance of 
$\log\epsilon_{\element{Fe},\sun}
=7.50$~\footnote{$\log\epsilon_{\element{Fe}}\equiv
\log(n_{\element{Fe}}/n_{\element{H}})+12.$}. 
The chemical composition of the other models was scaled 
with respect to solar according to their LTE iron abundance
and assuming an $\alpha$-enhancement
of $[\alpha/\mathrm{Fe}]=0.4$ dex for the two metal-poor stars
in the sample.

Along with the {\sc marcs} solar atmosphere, we considered
two additional models for the Sun:
the semi-empirical Holweger-M{\"u}ller ({\sc hm}) model 
(Holweger \& M{\"u}ller \cite{holweger}) and the quiet-Sun
{\sc val-3c} model by Vernazza et al. (\cite{vernazza}).
In this way we intended to cover -- at least for the Sun -- 
typical model-to-model variations in the analysis.
The temperature stratifications for the three solar models 
are shown in Fig. \ref{marcsatm} together with models for 
the other three stars. 
The {\sc marcs} and {\sc hm} models have 
similar stratifications although the latter has slightly
warmer temperatures in the line forming regions.
The non-LTE {\sc val-3c} model departs from the other two solar models 
mainly by presenting a chromospheric temperature rise in its 
upper layers and reaching significantly lower temperatures 
at optical depths $-4\la\log(\tau_{5000})\la-1.5$. 
For a quantitative description of how structural differences 
arise between LTE and non-LTE models of the solar photosphere 
we refer to the work by Rutten \& Kostik (\cite{rutten}) and 
Rutten (\cite{rutten2}). 

\begin{table*}[thp]
	\centering
	\caption{\element{Fe} abundance corrections 
	$\Delta{\log\epsilon}_\mathrm{Fe}
       	= ({\log\epsilon}_\mathrm{Fe})_\mathrm{\,non-LTE}-({\log\epsilon}_\mathrm{Fe})_\mathrm{\,LTE}$
       	for our sample stars.
	The results for calculations with and without line-blocking
	are shown, for two different values of the efficiency $S_\mathrm{H}$ 
	of \ion{H}{i} collisions. 
	The reported values of $\Delta{\log\epsilon}_\mathrm{Fe}$
	are averages over weak \ion{Fe}{i} lines in the visible and near UV with equivalent
	widths between $5$~m{\AA} and $100$~m{\AA}. 
	The standard deviation is given  as a measure of the line-to-line scatter
	in the non-LTE abundance corrections. No direct comparison with
	observations has here been undertaken.
	}
	\begin{tabular*}{0.9\hsize}{lllll}
           \hline
	   \hline
	   \noalign{\smallskip}
	   \multicolumn{5}{c}{ {\sc \element{Fe} non-LTE Abundance Corrections from \ion{Fe}{i} Lines} }\\
	   \noalign{\smallskip}
	   \hline
	   \noalign{\smallskip}
	   \hspace{3cm}  & \multicolumn{2}{l}{\hspace{1.5cm}$S_{\mathrm{H}}=0.001$}&
	   \multicolumn{2}{l}{\hspace{1.8cm}$S_{\mathrm{H}}=1$}\\
	   \noalign{\smallskip}
	   Model&
	   No blocking \hspace{1cm} &
	   With blocking   \hspace{1.5cm} &
	   No blocking \hspace{1cm} &
	   With blocking   \hspace{1cm} \\
	   \noalign{\smallskip}
 	   \hline
           \noalign{\smallskip}
           \noalign{\smallskip}
 	   \object{Sun} \,({\sc marcs}) & $0.24 \pm 0.02  $ & $0.08 \pm 0.02$ & $0.12 \pm 0.01$   & $0.03  \pm 0.01$ \\
 	   \object{Sun} \,({\sc hm})	& $0.22 \pm 0.02  $ & $0.07 \pm 0.02$ & $0.13 \pm 0.02$   & $0.04  \pm 0.01$ \\
 	   \object{Sun} \,({\sc val-3c})& $0.25 \pm 0.04  $ & $0.12 \pm 0.03$ & $0.17 \pm 0.03$   & $0.07  \pm 0.02$  \\
 	   \object{Procyon}		& $0.23 \pm 0.03  $ & $0.09 \pm 0.02$ & $0.13 \pm 0.02$  & $0.04 \pm 0.02$ \\
 	   \object{HD~140283}		& $0.58 \pm 0.05  $ & $0.56 \pm 0.05$ & $0.32 \pm 0.02$  & $0.30 \pm 0.02$ \\
 	   \object{G64-12}		& $0.69 \pm 0.06 $  & $0.68 \pm 0.06$ & $0.48 \pm 0.03$  & $0.47 \pm 0.03$ \\
           \noalign{\smallskip}
           \noalign{\smallskip}
	   \hline
       	\end{tabular*}
        \label{tabnlteco1}
\end{table*}

\subsection{Radiative transfer calculations}
 We computed the emerging \element{Fe} spectrum 
for each of the model atmospheres by solving the statistical
equilibrium and radiative transfer equations with
the non-LTE code {\sc multi} (Carlsson \cite{carlsson}), version 2.2.
We introduced however some modifications  concerning
the treatment of background opacities.
{\sc multi}'s standard background opacities are
based on the Uppsala package (Gustafsson \cite{gustafsson})
which comprises continuous opacities only.
Several \element{Fe} bound-free extinction edges
though are located in the ultraviolet where a large number
of atomic lines from metals contribute 
to the UV ``line haze'' typically encountered 
in the spectra of cool stars.
Accurate evaluation of the photo-ionisation rates 
in correspondence with these bound-free transitions
requires that we take into account the effects of 
line-blocking associated with the UV line haze.

Some previous attempts have been made to include the effects of line-blocking
in non-LTE calculations in {\sc multi}. 
In their study of non-LTE formation of the solar spectrum of
alkali  Bruls et al. (\cite{bruls}) reproduced the contribution
of atomic lines to the background opacities by multiplying
the total \element[-]{H} and continuous metal opacities
with wavelength dependent factors. 
Their factors were determined empirically although only for the Sun
by fitting the observed solar continuum with the emerging 
continuum intensity computed with the {\sc val-3c} model
solar atmosphere and assuming they are independent of depth.
Andretta et al. (\cite{andretta}) introduced background atomic
and molecular lines to the standard  {\sc multi}'s
opacities in the form of an opacity table provided
with the particular model atmosphere they used.
The opacity dependence upon wavelength was given 
in their table at each depth point of the model atmosphere 
with a resolution of $2$~{\AA}. The opacity data they used 
were fixed however and model-specific.
More recently Bayazitov (\cite{bayazitov}) investigated the
effects of line-blocking on non-LTE \ion{Fe}{i} line formation
by adding the contribution of about $18\,000$ spectral lines
to {\sc multi}'s continuous opacities. His work however was restricted to
stars with metallicities close to solar and relied on a much simpler model 
\ion{Fe}{i} atom with $99$ levels plus continuum.

In our study, in order to include the effects of
line-blocking in {\sc multi} we proceeded 
by sampling metal line opacities for about $9\,000$
wavelength points in the spectral region between $1\,000$~{\AA}
and $20\,000$~{\AA} and adding them to the standard
background continuous opacities. 
In order not to count the contribution of \element{Fe} line
opacities twice however, metal line opacities were only added 
to the background when computing intensities for the photo-ionisation rates and not for 
\ion{Fe}{i} and \ion{Fe}{ii} line profiles.
Molecular line opacities were not considered in our line formation
calculations which however is a reasonable approach for our stars where the atomic
contribution dominates overwhelmingly.
We performed simple bench-mark test LTE radiative transfer calculations
with background line opacities sampled  between $9\,000$ and $70\,000$ wavelength 
points covering the same wavelength region. 
The tests indicate that photo-ionisation rates computed with
our choice of opacity sampling for $9\,000$ wavelengths are 
sufficiently accurate for our purposes.
In addition, we performed non-LTE test calculations sampling the opacities 
at different wavelengths and found that the overall effect on the final 
\element{Fe} non-LTE abundance corrections is less then $0.01$~dex.

 For the calculations presented in this paper 
the background metal line opacities were computed in detail 
by our modified version of {\sc multi},
consistently with the particular model atmosphere 
given as input, as a function of temperature, gas- 
and electron-pressure at each depth point. 
 For this purpose we used the opacity package from the 
{\sc os-marcs} code (Gustafsson et al. \cite{gustafsson2})
taking into account the specific chemical composition 
of the given model stellar atmospheres. 
Level populations and ionisation equilibria
for the various metals -- including \element{Fe} --
were computed assuming LTE. 
This inconsistency for \element{Fe} is however of minor importance
and may well be balanced by missing line-blocking for other species.

The non-LTE abundance corrections were estimated by computing
both the LTE and non-LTE cases for several different abundances,
from which the non-LTE 
\element{Fe} abundance that reproduced the LTE equivalent widths
were interpolated. For each set
of calculations with a given model atmosphere
we selected the \ion{Fe}{i} spectral lines in 
our model atom falling in the visible and in the near UV,
($3\,000$~{\AA} $\leq \lambda \leq  10\,000$~{\AA}),
and with equivalent widths $W_{\lambda}$ between $5$~m{\AA}
and $100$~m{\AA}. 

\section{Results}
\begin{figure*}[tph]
	\centering
   	\resizebox{\hsize}{!}{
		\includegraphics{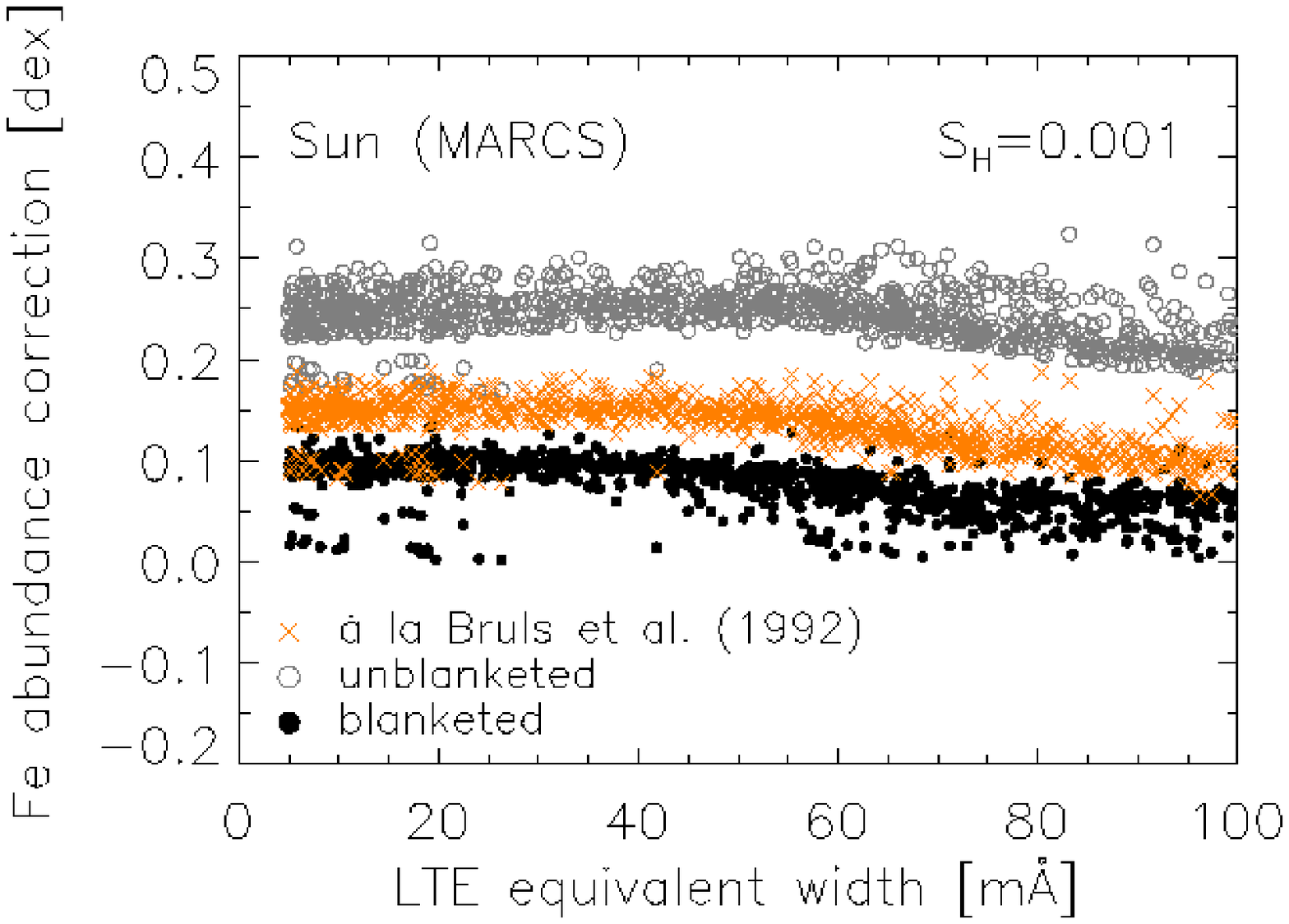}
   		\includegraphics{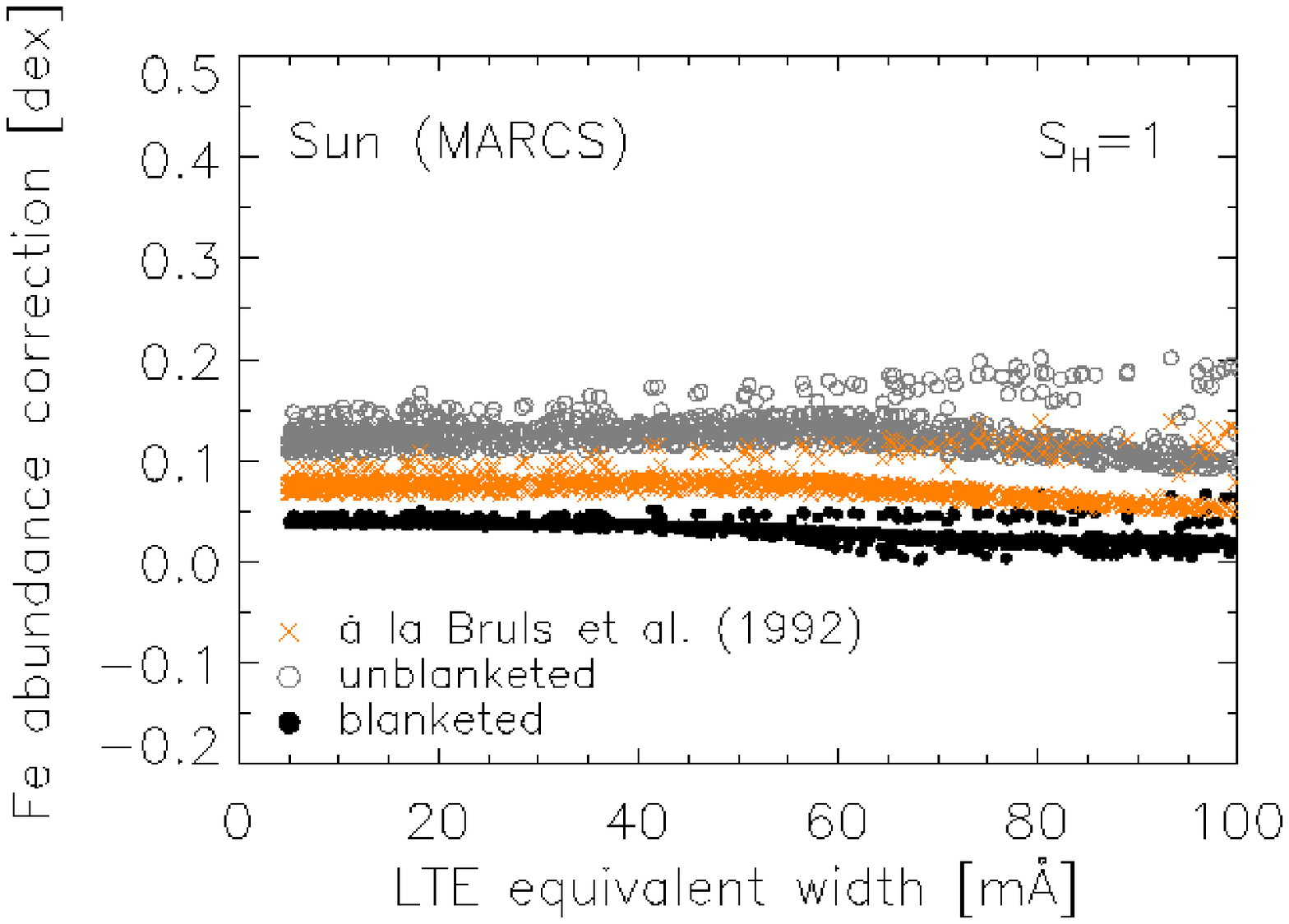} }
   	\resizebox{\hsize}{!}{
		\includegraphics{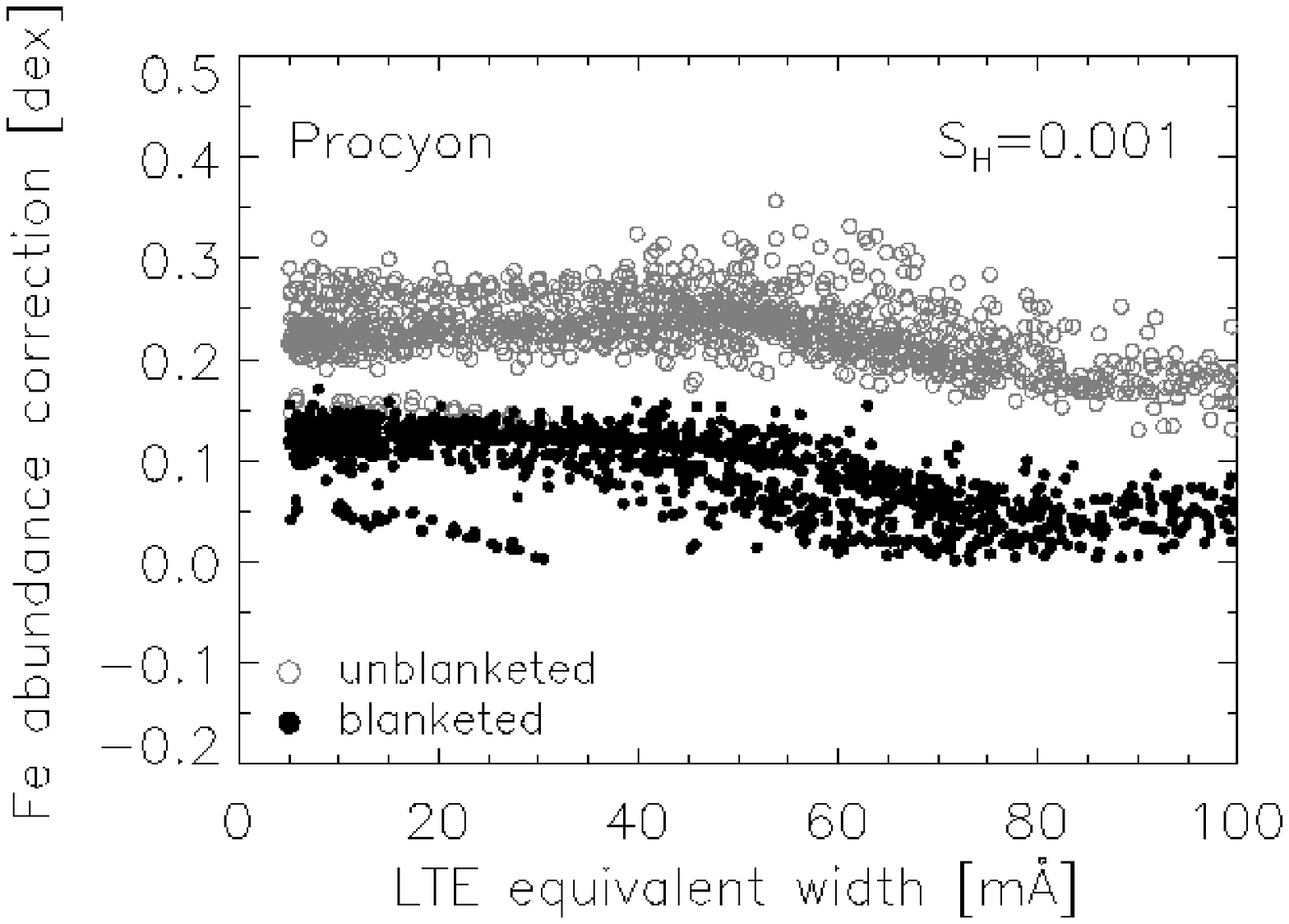}
   		\includegraphics{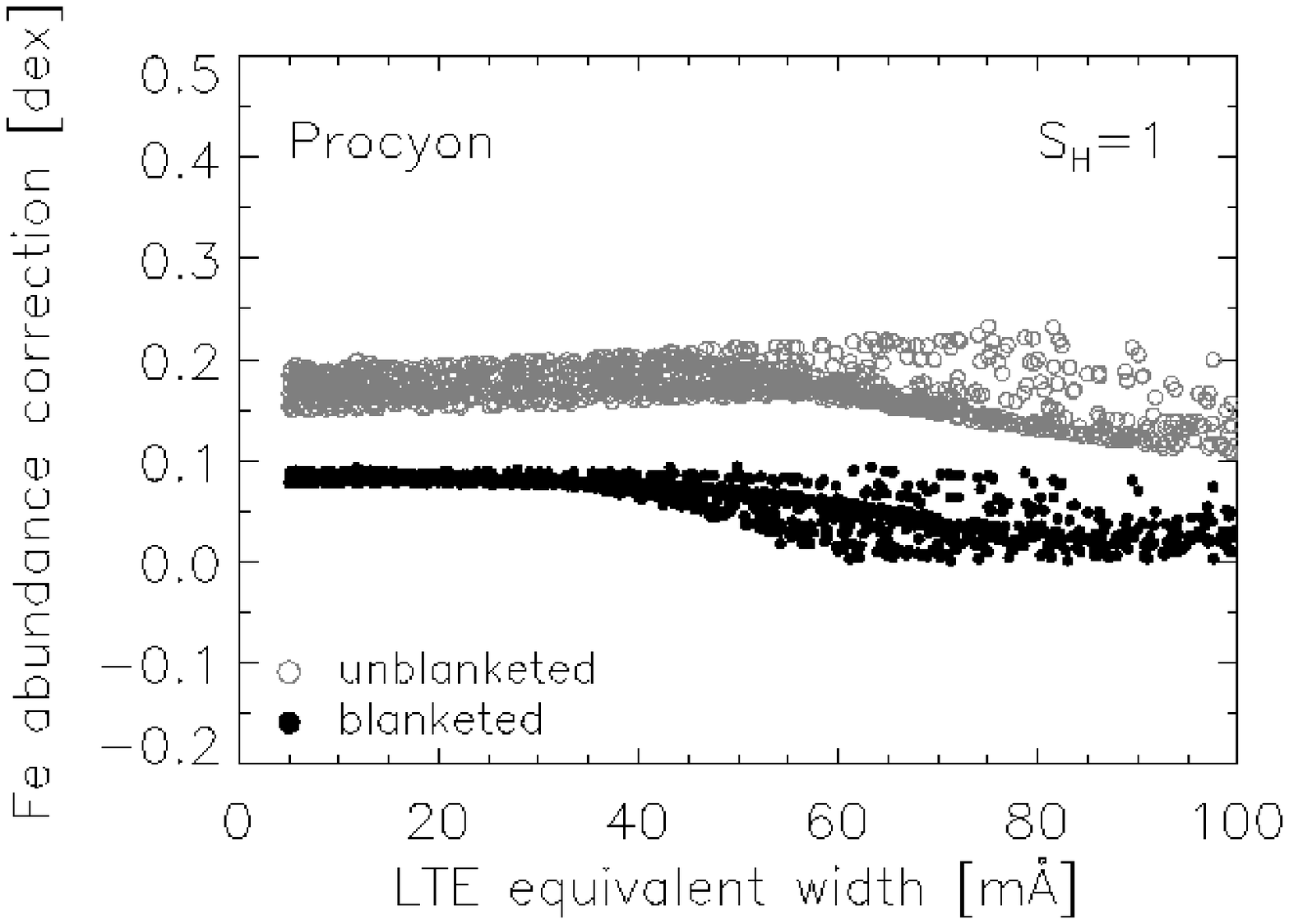} }
     	\caption{Non-LTE \element{Fe} abundance corrections 
       		for the \object{Sun}'s (Top) and \object{Procyon}'s (Bottom)
		{\sc marcs} model atmospheres. 
       		The results are presented for two values for the efficiency 	
	       	$S_{\mathrm{H}}$ of \ion{H}{i} collisions:  
       		$S_{\mathrm{H}}=0.001$ (Left) and $S_{\mathrm{H}}=1$ (Right).
       		Empty circles refer to the calculations without 
       		background line opacities included, filled circles 
       		to the calculations with background line opacities. 
       		For the \object{Sun} the corrections computed including 
       		background line opacities \textit{{\`a} la} Bruls et al. (1992) 
       		are also plotted (crosses).}
	\label{nltecsolar}
\end{figure*} 
 
\begin{figure*}[thp]
   	\centering
   	\resizebox{\hsize}{!}{
		\includegraphics{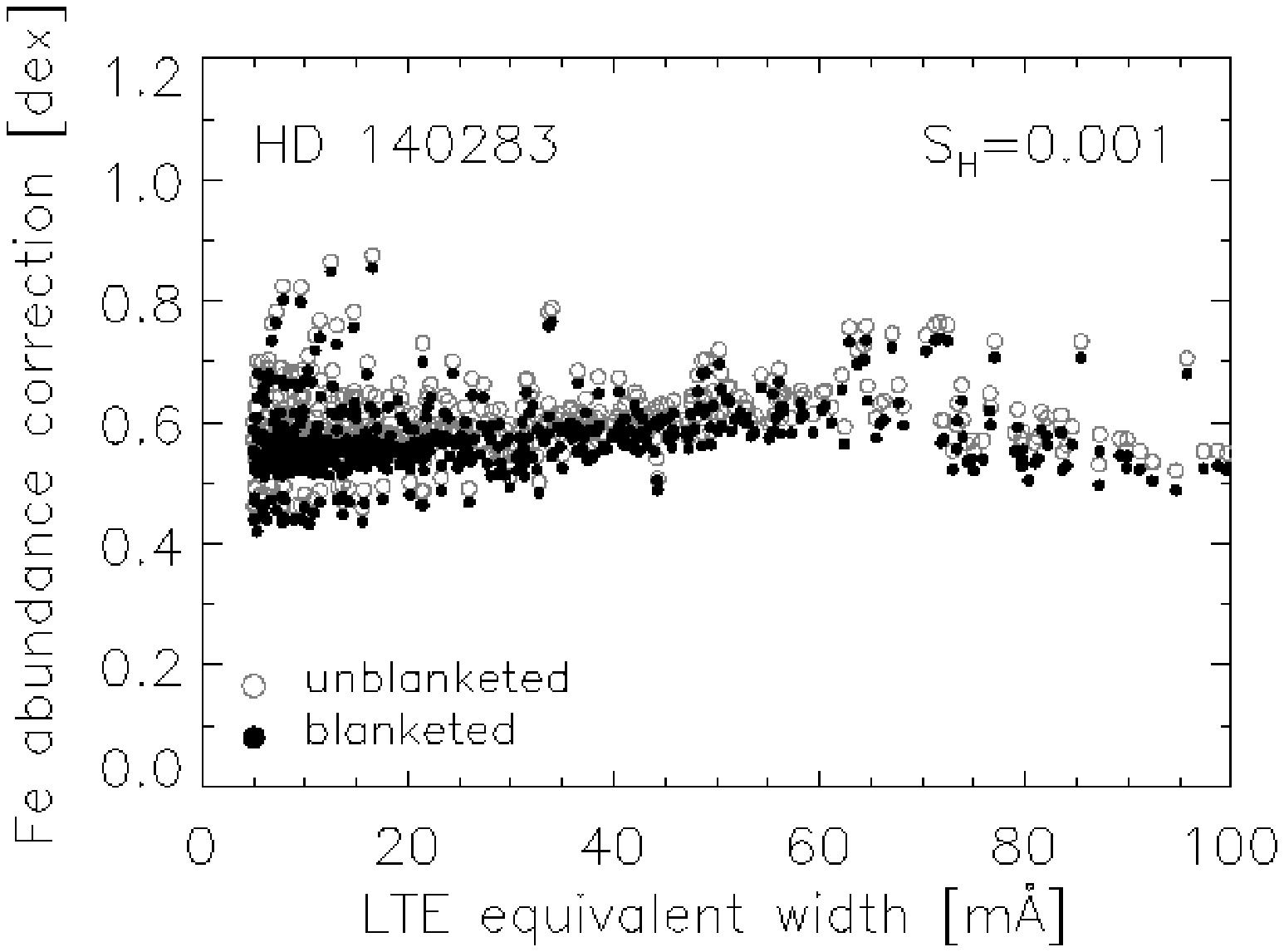}
   		\includegraphics{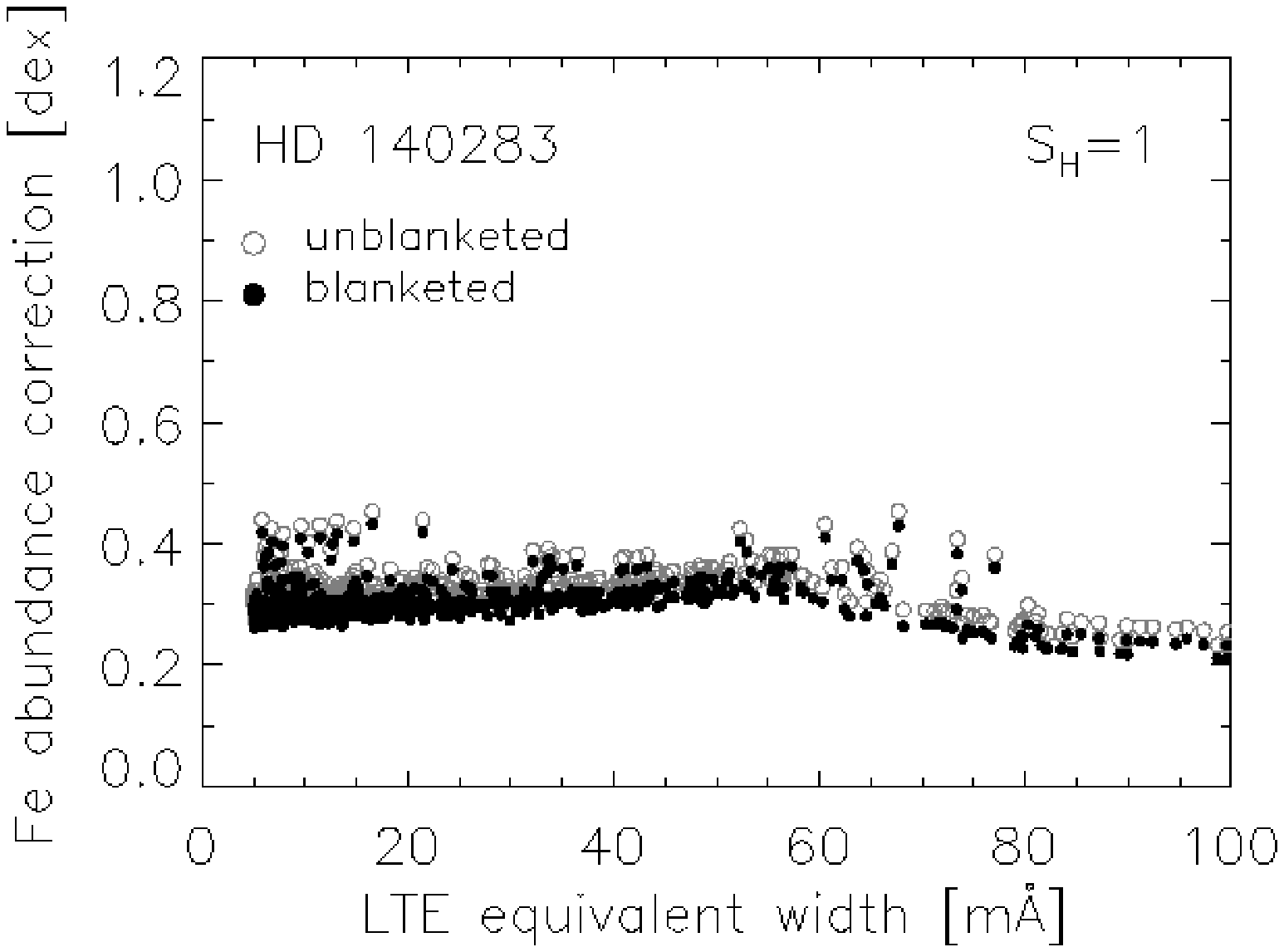} }
   	\resizebox{\hsize}{!}{
		\includegraphics{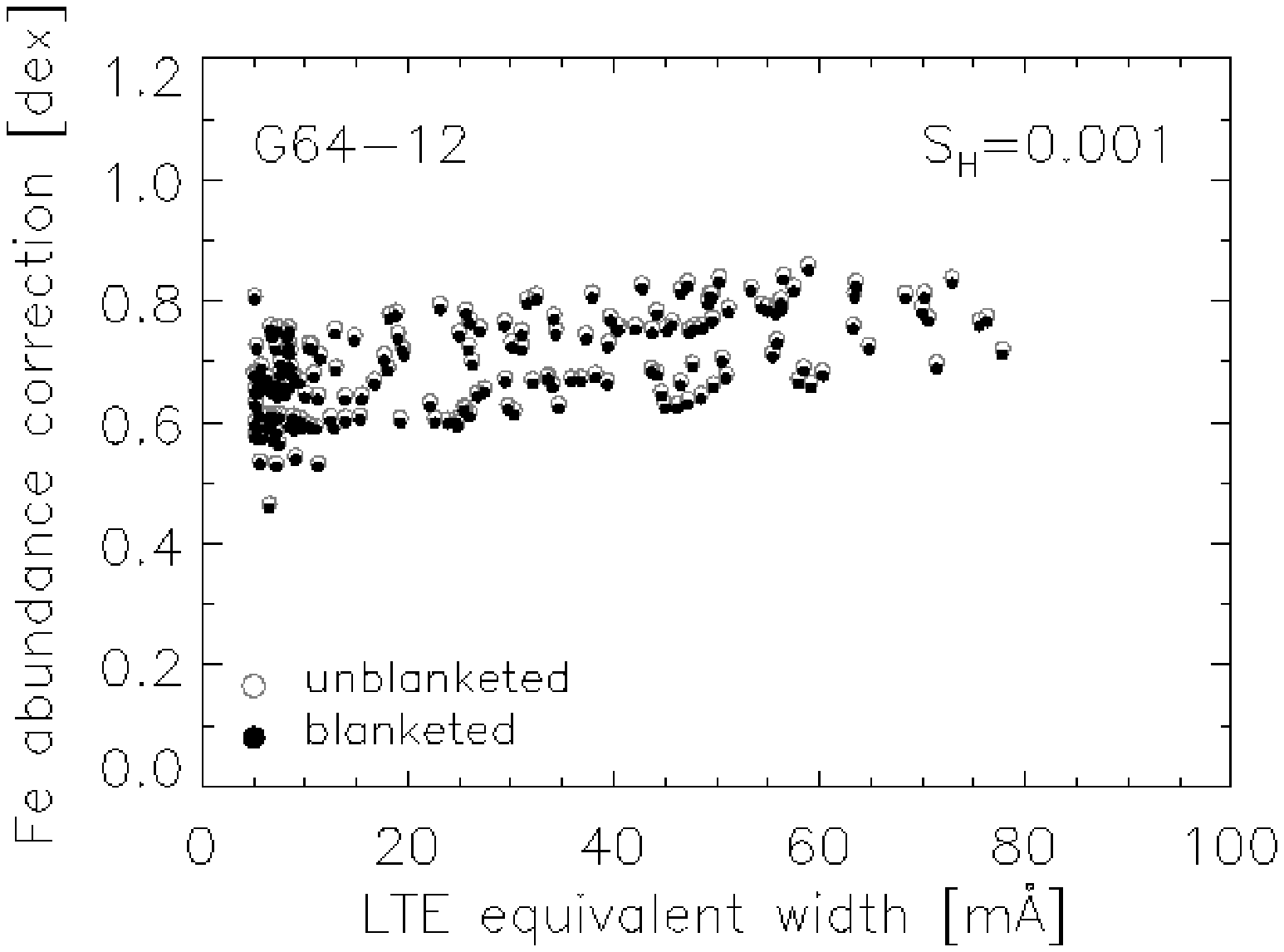}
   		\includegraphics{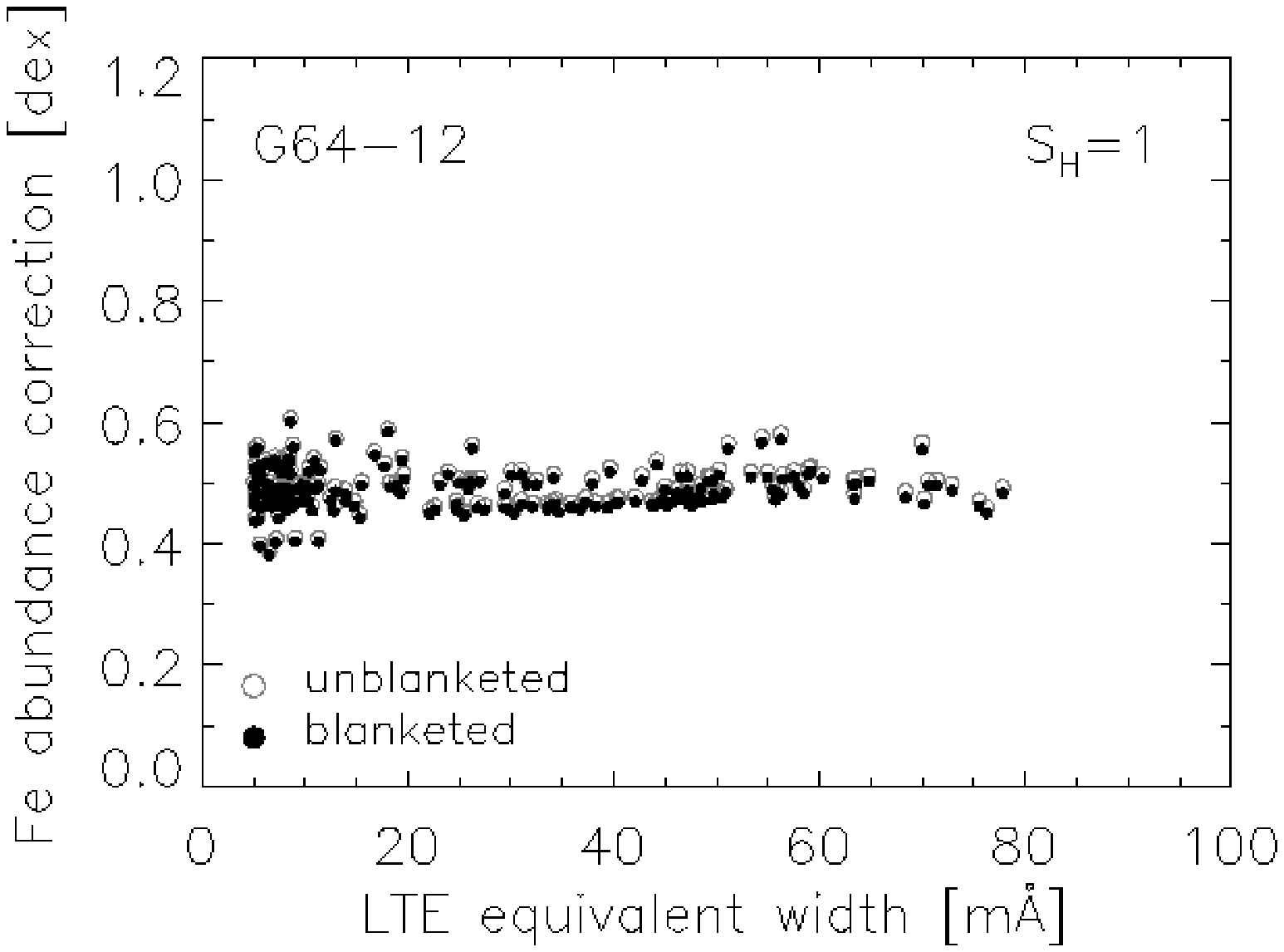}  }
     	\caption{Non-LTE \element{Fe} abundance corrections 
       		for the metal-poor stars \object{HD~140283}'s (Top) 
		and \object{G64-12}'s (Bottom)
       		{\sc marcs} model atmospheres. 
       		The results are presented for two values for the efficiency 
       		$S_{\mathrm{H}}$ of \ion{H}{i} collisions:  
       		$S_{\mathrm{H}}=0.001$ (Left) and $S_{\mathrm{H}}=1$ (Right).
       		Empty circles refer to the calculations without 
       		background line opacities included, filled circles 
       		to the calculations with background line opacities.}
	\label{nltecmetalpoor}
\end{figure*}

Figure \ref{nltecsolar} shows the non-LTE corrections to the
iron abundance derived for the {\sc marcs} model solar atmosphere
for our selection of \ion{Fe}{i} lines; the behaviour for the other
two solar models is similar as summarised in Table \ref{tabnlteco1}. 
Results are presented
for the two different efficiencies of Drawinian collisions adopted
in this study, $S_{\mathrm{H}}=0.001$ and $S_{\mathrm{H}}=1$, and for both 
the case including line-blocking and the one excluding it.

As a consequence of over-ionisation the overall line opacity of weak neutral 
iron lines decreases and the lines form deeper into the atmosphere in non-LTE 
than with the assumption of LTE, as already shown by Athay \& Lites (\cite{athaylites}). 
The equivalent widths of these lines thus decrease and 
therefore a positive correction has to be applied to the LTE iron abundance
derived from \ion{Fe}{i} lines.
As expected, the non-LTE effects are more severe in models with a steeper temperature
structure in the line forming layers as the {\sc val-3c} model
(see discussion in Asplund \cite{asplund2}).

Let us first examine the abundance corrections in the case of low-efficiency 
\ion{H}{i} collisions.
When background line opacities are not included in the calculations
we estimate a correction to the iron abundance of about $0.24$~dex for the weak 
($W_\lambda \la 50$~m{\AA}) solar \ion{Fe}{i} lines in our sample 
with little dependency on the equivalent width $W_\lambda$. 
Weak lines form generally deep in the stellar atmosphere where the source 
function $S_\nu$ is given by the local Planck function 
$B_\nu(T)$ and the  non-LTE equivalent width reflects mainly 
\ion{Fe}{i} depopulation in the region of line formation. 
Stronger lines ($W_\lambda \ga 50$~m{\AA}) on the contrary form further out 
where the source function at line centre typically falls below the Planck function
compensating in part for the effects of \ion{Fe}{i} depletion on
the line strength (Saxner \cite{saxner}).

With the inclusion of sampled background line opacities in the
calculations the mean intensity
field $J_\nu$ at a given depth decreases and so do the radiative ionisation
rates which are given by $R_\mathrm{ion}=\int 4\pi\frac{\sigma_\nu}{h\nu}J_\nu d\nu$
where $\sigma_\nu$ indicates the photo-ionisation cross-section.
A lower ratio between radiative ionisations and recombinations
and therefore a weaker over-ionisation are expected.
Consequently the non-LTE \element{Fe} abundance corrections also 
decrease and fall to values of about $0.08$~dex (average over lines
with $W_\lambda$ between $5$ and $100$~m{\AA}).

We investigated as well the effects of line-blocking when multiplication
factors  were applied to the continuum opacities to simulate the UV line
haze according to the recipe of Bruls et al. (\cite{bruls}). 
The results indicate that 
over-ionisation is indeed reduced in this case but by a smaller amount:
the predicted \ion{Fe}{i} non-LTE abundance corrections
remain about $0.05$~dex higher than in our calculations with line-blocking.
The reason for the difference is that the technique by
Bruls et al. (\cite{bruls}) results in overall lower mean background opacities 
than in the case of the full opacity sampling method.

When the rates of collisions with neutral hydrogen are increased 
($S_\mathrm{H}=1$) a lower degree of over-ionisation 
is achieved as expected due to increased thermalisation. 
The equivalent widths of non-LTE weak neutral iron lines therefore increase and
the overall non-LTE \element{Fe} abundance correction is reduced as illustrated
in Fig. \ref{nltecsolar} and in Table \ref{tabnlteco1}. 
Because of the overall lower non-LTE abundance corrections, the impact
of background line opacities  is less pronounced with a difference of about $0.1$~dex 
between the calculations including and excluding line-blocking at solar metallicity.

Concerning \object{Procyon} (Fig. \ref{nltecsolar}), departures 
of \element{Fe} abundance from LTE calculations are similar
to the solar case with and without line-blocking
(Table \ref{tabnlteco1}). Because of its higher $T_\mathrm{eff}$
and lower $\log g$ \object{Procyon} shows slightly larger non-LTE
iron abundance corrections than the \object{Sun}
for weak \ion{Fe}{i} lines.
Multiplication factors from Bruls et al. (\cite{bruls}) 
were calibrated specifically for the solar atmosphere and therefore were
not applied to simulate line-blocking in Procyon's atmosphere.

\begin{table*}[thp]
	\centering
	\caption{\element{Fe} non-LTE abundance corrections 
      	for our sample stars estimated from weak \ion{Fe}{ii} lines in the visible and near UV.
	Averages are computed over weak \ion{Fe}{ii} lines in the visible and near UV with equivalent
	widths between $5$~m{\AA} and $100$~m{\AA}. The standard deviation  
	is given as a measure of the line-to-line scatter in the non-LTE abundance corrections. } 
\begin{tabular*}{0.9\hsize}{lllll}
           \hline
	   \hline
	   \noalign{\smallskip}
	   \multicolumn{5}{c}{ {\sc \element{Fe} non-LTE  Abundance Corrections from \ion{Fe}{ii} Lines} }\\
	   \noalign{\smallskip}
	   \hline
	   \noalign{\smallskip}
	   \hspace{3cm}  & \multicolumn{2}{l}{\hspace{1.5cm}$S_{\mathrm{H}}=0.001$}&
	   \multicolumn{2}{l}{\hspace{1.8cm}$S_{\mathrm{H}}=1$}\\
	   \noalign{\smallskip}
	   Model&
	   No blocking \hspace{1cm} &
	   With blocking   \hspace{1.5cm} &
	   No blocking \hspace{1cm} &
	   With blocking   \hspace{1cm} \\
	   \noalign{\smallskip}
 	   \hline
           \noalign{\smallskip}
           \noalign{\smallskip}
 	   \object{Sun} \,({\sc marcs}) & $0.02 \pm 0.01  $ & $0.01 \pm 0.01$ & $0.00 \pm 0.00$   & $0.00  \pm 0.00$ \\
 	   \object{Sun} \,({\sc hm})	& $0.02 \pm 0.01  $ & $0.01 \pm 0.01$ & $0.00 \pm 0.00$   & $0.00  \pm 0.00$ \\
 	   \object{Sun} \,({\sc val-3c})& $0.00 \pm 0.00  $ & $0.00 \pm 0.00$ & $0.00 \pm 0.00$   & $0.00  \pm 0.00$  \\
 	   \object{Procyon}		& $0.02 \pm 0.01  $ & $0.02 \pm 0.01$ & $0.01 \pm 0.00$   & $0.01  \pm 0.00$ \\
 	   \object{HD~140283}		& $0.24 \pm 0.13  $ & $0.24 \pm 0.13$ & $0.01  \pm 0.01$  & $0.01  \pm 0.01$ \\
 	   \object{G64-12}		& $0.17 \pm 0.08 $  & $0.17 \pm 0.08$ & $0.05  \pm 0.06$  & $0.05  \pm 0.06$ \\
           \noalign{\smallskip}
           \noalign{\smallskip}
	   \hline
       \end{tabular*}
         \label{tabnlteco2}
\end{table*}

\begin{figure*}[thp]
	\centering
	\resizebox{\hsize}{!}{
		\includegraphics{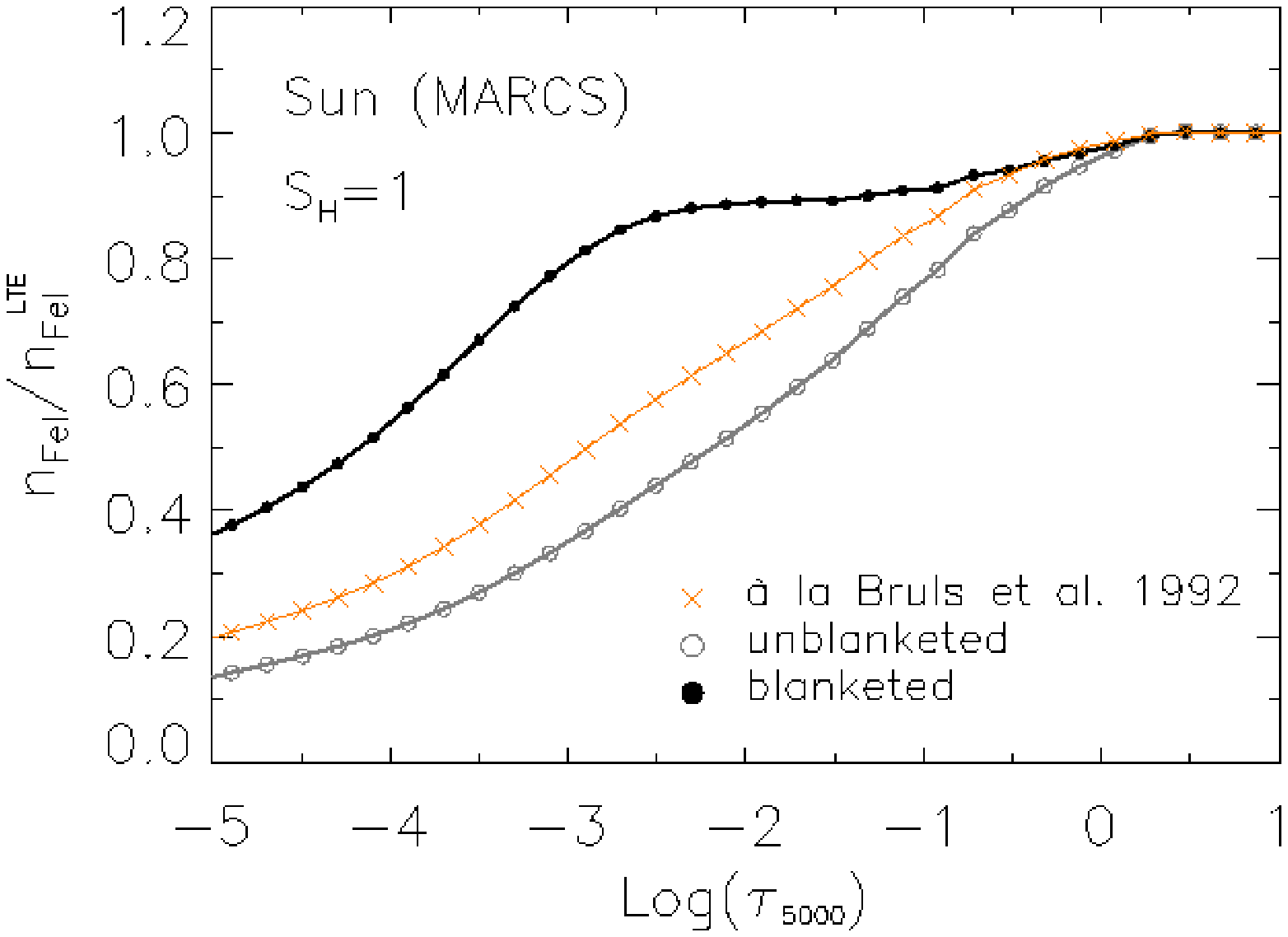}
   		\includegraphics{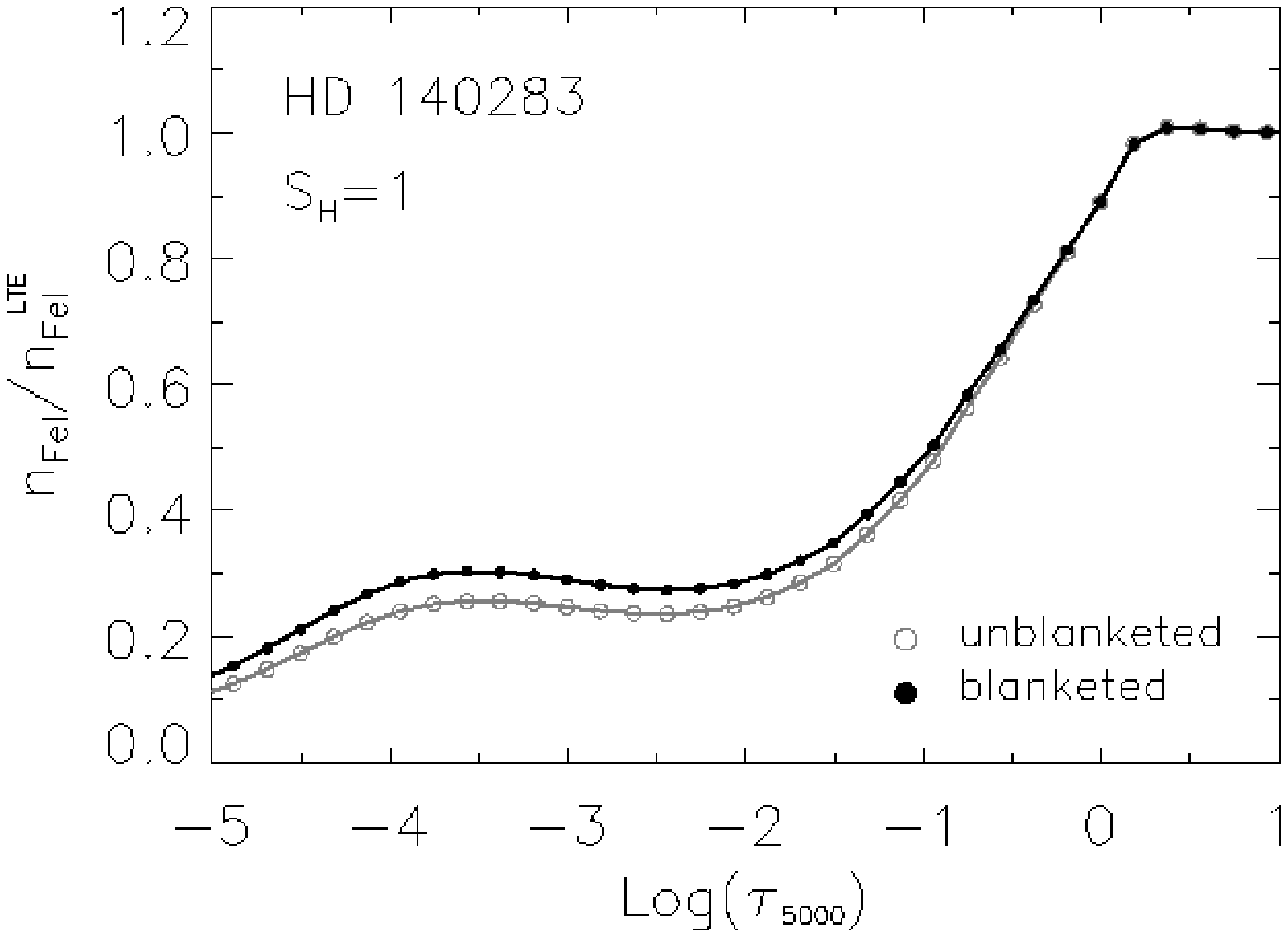} }
   	\caption{Departure of total \ion{Fe}{i} number density $n_\mathrm{\,\ion{Fe}{i}}$
   	from LTE versus standard optical depth at $\lambda=5000$~{\AA}  for the
	\object{Sun} (Left) and for \object{HD~140283} 	(Right). 
	non-LTE departures are shown for the case of fully efficient Drawinian
   	collisions and for different treatments of line-blocking.}
     	\label{popfei}
\end{figure*}

In contrast to the solar metallicity case, 
the inclusion of the line haze at low metallicities produces 
minor changes in terms of mean intensity $J_\nu$ due to the overall low 
contribution of lines to the total opacities. 
The exclusion of background line opacities still produces systematically 
larger mean intensities at all depth and therefore generally increases
photo-ionisation rates and non-LTE \element{Fe} abundance corrections
but only very slightly so.
The difference between the results of calculations with and without line-blocking is 
typically only $0.02$~dex for both \object{HD~140283} and \object{G64-12}.

Figure \ref{popfei} illustrates the effects of line-blocking on 
the departure from LTE of total \ion{Fe}{i} number density as a function
of depth for the {\sc marcs} model atmospheres of the Sun and the metal-poor
\object{HD~140283}; the departure coefficients for individual
\ion{Fe}{i} levels are all similar and considerable in the latter case. 
At solar metallicity the inclusion of line-blocking
significantly reduces the departure of the \ion{Fe}{i} population
from LTE in the region where most weak lines form.
On the contrary at low metallicity 
calculations with and without line-blocking produce very similar and significant
non-LTE signatures for a large portion of the line forming region in the atmosphere.
Due to the overall low contribution of the line haze to the opacities 
at low metallicities, the mean intensities are typically larger than at solar 
metallicity. 
Consequently photo-ionisation rates are higher and departures from LTE
more pronounced in metal-poor stars.
In fact the departure of \ion{Fe}{i} population from LTE in metal-poor stars 
is controlled primarily by the efficiency of the \ion{H}{i} collisions. 
Changing the scaling factor $S_\mathrm{H}$ in Drawin's formula from $0.001$
to $1.0$ decreases the non-LTE abundance corrections by $0.2-0.3$~dex for
the metal-poor stars studied here.

It is clear that taking account of line-blocking is important
for the \ion{Fe}{i} non-LTE line formation at solar metallicity 
but significantly less so at low metallicity. Instead, the dominant
source of error in metal-poor stars is the treatment of inelastic 
\ion{H}{i} collisions, with the non-LTE abundance corrections differing
by $0.2-0.3$\,dex when using  $S_\mathrm{H} = 1$ and
 $S_\mathrm{H} = 0.001$ (Table \ref{tabnlteco1}).
While our results for the metal-poor case with $S_\mathrm{H}=0.001$
are very similar to the findings of Korn et al. (\cite{korn}) with no
hydrogen collisions,
we note that our Fe non-LTE abundance corrections with $S_\mathrm{H} = 1$
are still significantly larger than in their calculations. 
The main reason for this is that in addition to using 
even more efficient \ion{H}{i} collisions ($S_\mathrm{H} = 3$), they also
more importantly
enforce ad-hoc thermalisation of the uppermost levels of  \ion{Fe}{i} 
with the ground level of \ion{Fe}{ii}. Since the latter is the dominant
\element{Fe} state and is well described by LTE, all \ion{Fe}{i} levels
are steered closer to the LTE solution by bound-bound collisions
within the \ion{Fe}{i} system. 
In fact test calculations performed with their model atom 
for \object{HD~140283} and with $S_\mathrm{H}=1$ but no enforced 
thermalisation of highly excited \ion{Fe}{i} levels produce similarly large
non-LTE effects, i.e. $\Delta{\log\epsilon}_\mathrm{Fe}\approx0.45$~dex 
compared with our value of  $0.3$~dex (Korn 2005, private communication).
Our non-LTE abundance corrections are also much larger than the corresponding
computations of Gratton et al. (\cite{gratton}), which is not
surprising given their very efficient \ion{H}{i} collisions ($S_\mathrm{H} \approx 30$)
and their use of considerably smaller photo-ionisation cross-sections
than those of Bautista et al. (\cite{bautista}).

Overall we find slightly larger non-LTE corrections than 
Th{\'e}venin \& Idiart (\cite{thevenin}) even when we consider fully efficient 
\ion{H}{i} collisions ($S_\mathrm{H}=1$) as it was actually used in their work.
On the other hand the approach of Th{\'e}venin \& Idiart (\cite{thevenin})
is not identical to ours. 
Their \ion{Fe}{i} and  \ion{Fe}{ii} systems were not merged and
their analysis relied on a different suite of model atmospheres.
As mentioned before our treatment of line-blocking  differs 
from theirs as they used multiplication factors to account for the background line haze
in a similar way as Bruls et al. (\cite{bruls}) did for the solar case.

Finally concerning \ion{Fe}{ii} lines, Table \ref{tabnlteco2} shows the 
\element{Fe} non-LTE abundance corrections estimated for weak singly ionised iron
lines in the visible and near UV. Contrary to what is often assumed
\ion{Fe}{ii} lines are not exempt from non-LTE effects. Departures of \ion{Fe}{ii} lines
from LTE can be substantial especially at low metallicity. In particular,
in the case of low \ion{H}{i} collisions \element{Fe} non-LTE abundance corrections
of about $0.2$~dex are found for metal-poor stars. Similar corrections 
for \object{HD~140283} are also reported by Shchukina et al. (\cite{shchukina2}).
For fully efficient \ion{H}{i} collisions corrections are however much lower although
not completely vanishing at low metallicity.
We note that the net \element{Fe} non-LTE effect (i.e. 
$\Delta{\log\epsilon}_\mathrm{\ion{Fe}{i}}-\Delta{\log\epsilon}_\mathrm{\ion{Fe}{ii}}$)
is quite similar for $S_H=0.001$ and $S_H=1$. It may therefore be difficult
to empirically discriminate between different choices of $S_H$ on the basis
of observations alone.

\section{Conclusions}
We have investigated the effects of line-blocking on non-LTE \element{Fe}
abundances derived from \ion{Fe}{i} lines for a selection of four representative
late-type stars.
We found the effects of line-blocking to be significant at solar metallicities.
In particular calculations including background line opacities  result in non-LTE  \element{Fe}
abundance corrections $0.1-0.15$~dex higher than in
calculations excluding them, depending on the efficiency of \ion{H}{i} collisions 
\textit{{\`a} la} Drawin ($S_\mathrm{H}=1$ and $S_\mathrm{H}=0.001$ respectively).
On the other hand line-blocking has a much smaller impact at low metallicities. 
Absolute non-LTE \element{Fe} abundance corrections are 
in this case significant and sensitive mainly to the strength of \ion{H}{i} collisions and vary only by 
$0.01-0.02$~dex between calculations with and without line-blocking.
At the present time the main uncertainty in non-LTE line formation calculations in metal-poor stars
is still the treatment of the poorly known inelastic \ion{H}{i} collisions
and whether or not thermalisation of the highly excited \ion{Fe}{i} levels
should be applied.

\begin{acknowledgements}
RC and MA have benefitted from travel support from {\sc STINT} 
(The Swedish Foundation for International Cooperation in Research 
and Higher Education). MA acknowledges
generous financial support from the Australian Research Council.

The authors would also like to thank A. E. Garc{\'\i}a P{\'e}rez
for help during the early stages of the project, M. Carlsson
for sharing his expertise on numerical radiative transfer and non-LTE line formation,
P. Barklem and A. Korn for fruitful discussions on 
the role of \ion{H}{i} collisions and B. Gustafsson for
valuable suggestions to the manuscript.

\end{acknowledgements}

\end{document}